# Acoustic Probing of the Jamming Transition in an Unconsolidated Granular Medium

X. Jacob, V. Aleshin, V. Tournat

*Laboratoire d'Acoustique, UMR-CNRS 6613, Universite du Maine, Avenue Olivier Messiaen, 72085 Le Mans, France*

P. Leclaire,

*Laboratoire de Recherche en Mécanique et Acoustique, I.S.A.T - Université de Bourgogne, 49, rue Mademoiselle Bourgeois, 58000 Nevers, France*

W. Lauriks

*Laboratorium voor Akoestiek en Thermische Fysica, Katholieke Universiteit Leuven, Celestijnenlaan 200D, BE-3001 Heverlee, Belgium*

V. E. Gusev*

*Laboratoire de Physique de l'Etat Condensé, UMR-CNRS 6087, Universite du Maine, Avenue Olivier Messiaen, 72085 Le Mans, France*

Experiments with acoustic waves guided along the mechanically free surface of an unconsolidated granular packed structure provide information on the elasticity of granular media at very low pressures that are naturally controlled by the gravitational acceleration and the depth beneath the surface. Comparison of the determined dispersion relations for guided surface acoustic modes with a theoretical model reveals the dependencies of the elastic moduli of the granular medium on pressure. The experiments confirm recent theoretical predictions that relaxation of the disordered granular packing through non-affine motion leads to a peculiar scaling of shear rigidity with pressure near the jamming transition corresponding to zero pressure. Unexpectedly, and in disagreement with the most of the available theories, the bulk modulus depends on pressure in a very similar way to the shear modulus.

Transition from fluidity to rigidity in disordered media is a poorly understood phenomena in mechanics. This problem keeps attracting the attention of researchers studying its particular manifestations such as the transition to glassiness in liquids [1,2] and to jamming in colloidal dispersions or granular packings [2,3]. General interest in the unjamming transition in unconsolidated granular media is supported by its macroscopic manifestation in nature in the form of avalanches of sand and snow [4].

Recent theoretical investigations [1,3] as well as numerical experiments with packing of elastic grains [5-10] have predicted that in the vicinity of the jamming transition in the solid phase the elasticity of a disordered material is controlled not just by the elasticity of the individual contacts between the grains but also by material relaxation via non-affine motion of the grains. In particular, the dependence of the shear rigidity $G$ on pressure $p$ deviates from the law $G \propto p^{1/3}$ that follows from the Hertz-Mindlin theory of contacts between spheres [11,12]. The theory [3] indicates that the shear modulus $G$ should also be proportional to the excess coordination number of contacts $\delta z$ with respect to the critical isostatic number $z_c$ at the jamming transition, where the number of the contacts in the system is equal to the number of force balance equations [1]. This results, due to the relation $\delta z \propto p^{1/3}$ [3], in $G \propto p^{2/3}$, while the bulk elastic modulus $B$ does not contain additional scaling with pressure: $B \propto p^{1/3}$. The scaling exponent 2/3 had earlier been found numerically for random packings of frictionless particles at zero temperature interacting through the repulsive Hertzian potential [5,6]; the corresponding range of pressures $p$ evaluated numerically can be estimated for glass beads to be $100\ \text{Pa} \le p \le 10\ \text{MPa}$. These results have been confirmed for $p \ge 100\ \text{kPa}$ in Ref. [7,8]. Moreover very recently numerical experiments [9,10] have demonstrated that the scaling is not modified by friction at least for the 2D packing of spheres. The deviation from the scalings $G \propto p^{\alpha_G}$ (with $\alpha_G = 2/3$) and $B \propto p^{\alpha_B}$ (with $\alpha_B = 1/3$) is predicted [7,10] only at higher pressures $p \ge 10\ MPa$, where the condition $\delta z << z_c$ (i.e. the vicinity of the jamming transition) no longer holds.

To the best of our knowledge there has only been a single attempt to experimentally verify these theoretical simulations, using photoelastic disks near the 2D jamming transition [13], in which reasonable agreement with theory was found. However the features predicted for the packing of spheres have not been tested experimentally. The measurement of the dependence on pressure of the shear and longitudinal sound velocities $c_{S,L}$, which under the condition of negligible variations of density scale as the square root of the elastic moduli, is a classical method for the evaluation of the state of an unconsolidated granular medium [14-16]. That the behavior $c_S \propto G^{1/2} \propto p^{\alpha_S}$ with $\alpha_S = 1/3$ has not been revealed in most of the earlier experiments could be attributed to an important deviation from the jamming threshold pressure $p = 0$ (for example, $p \ge 2\ \text{MPa}$ in [16], $p \ge 5\ \text{MPa}$ in [9]). However, the exponent $\alpha_S = 1/3$ has not been documented even in experiments conducted at much lower pressures ($5\ \text{kPa} \le p \le 50\ \text{kPa}$)[17] although there are indications that $\alpha_S > 1/4$ holds over a limited range of pressures $5\ \text{kPa} \le p \le 20\ \text{kPa}$. In a very limited interval $50\ \text{kPa} \le p \le 100\ \text{kPa}$ the exponent $\alpha_S = 0.3$, which is closer to 1/3 than to 1/4, has been proposed to fit the experimental data in Refs. [14,15]. It could be concluded that there are indications of fast scaling ($\alpha_S > 1/4$) in some acoustic experiments, but this phenomena has never been observed over a wide enough interval of sufficiently low pressures for the precise evaluation of the scaling exponent.

To access the elastic moduli of granular media at much lower pressures than usual we use waveguide acoustic modes propagating along the mechanically free surface of an unconsolidated granular medium [18,19]. The localization of these modes near the surface is due to the bending of the acoustic rays towards the surface caused by an increase with depth of the rigidity of the gravity-loaded granular packing (the acoustic mirage effect[20]). Through experiments at ultrasonic frequencies ($500\,\text{Hz} \leq f \leq 5\,\text{kHz}$, an order of magnitude higher than in earlier experiments [21,22]), we are able to test near-surface layers at low pressures corresponding to the vicinity of the jamming transition. The lowest-order waveguide mode at its highest frequency $f = 3\,\text{kHz}$ propagates at a velocity ~15 m/s and penetrates only ~0.5 cm beneath the surface, probing the material at pressures less than 75 Pa. This is below the lowest pressure level accessed in numerical simulations [7,8]. However, even at the bottom of the experimental container (Fig. 1) at a depth of 20 cm, which can be accessed through the low-frequency higher-order modes, the pressure does not exceed 7.5 kPa, and is lower than in the most of the earlier experiments with bulk acoustic waves. The new experimental method proposed here to study such properties of granular materials consists in measuring at the surface of a granular layer the motion associated with the propagation of guided surface acoustic modes. The waves are excited using a 1 mm thick aluminium plate attached to a shaker and partly buried in the granular material (Fig. 1). The vertical surface velocity as a function of time and distance is measured with the help of a laser Doppler vibrometer with a surface focused beam (sensitivity 1mm/s/V within the frequency band 80 Hz – 20 kHz). The particle velocity is recorded every millimeter along the propagation direction from the source over a 300 mm distance. Typically, a maximum vertical velocity of $5.10^{-4}$ m/s at 1 kHz was observed close to the source. As the mechanical nonlinearity of a granular medium increases on approaching the jamming point, [6,17] we checked that the data are not influenced by the wave amplitude in the chosen range of excitation strengths. Using the measured signal in the time-space domain it is straightforward to perform the time and space double Fourier transform of the signals and to obtain frequency-wavenumber signals and plots of the dispersion curves for different guided acoustic modes. The experiments are performed on a granular material consisting of glass beads 150 $\mu$m in diameter in a large tank of dimensions 80 cm × 50 cm × 20 cm (see Fig. 1). The sample is prepared before each experiment as follows: the container is filled up with the grains and is shaken gently. Then acoustic excitations of different amplitudes and frequency ranges are applied for several hours until a stable response of the medium is obtained.

A typical experimentally obtained pattern in the ($\omega$, $k$)-domain is presented in Figure 2. The range 0<$k$<50-100 m$^{-1}$ is probably affected by the reflection of the waves from the container walls, but the short-wave part corresponds to the modes guided in the horizontal channel [18,19,22].

The dispersion relations in Fig. 2 are plotted using an intensity scale corresponding to the logarithm of the the normalized Fourier transform of the particle velocity. In order to further increase the contrast, threshold filtering was performed. The measurements were repeated a number of times with different excitation functions, namely short pulses with Gaussian spectra centered at different frequencies. For each dataset, maxima of the smoothed intensity function for different $k$-points were found. In 80% of the different experimental conditions these maxima were in good agreement with each other. We selected only these points and calculated the power-law approximation curves together with the standard deviation for each mode. In Fig. 2 the solid lines represent the averages ± standard deviations.

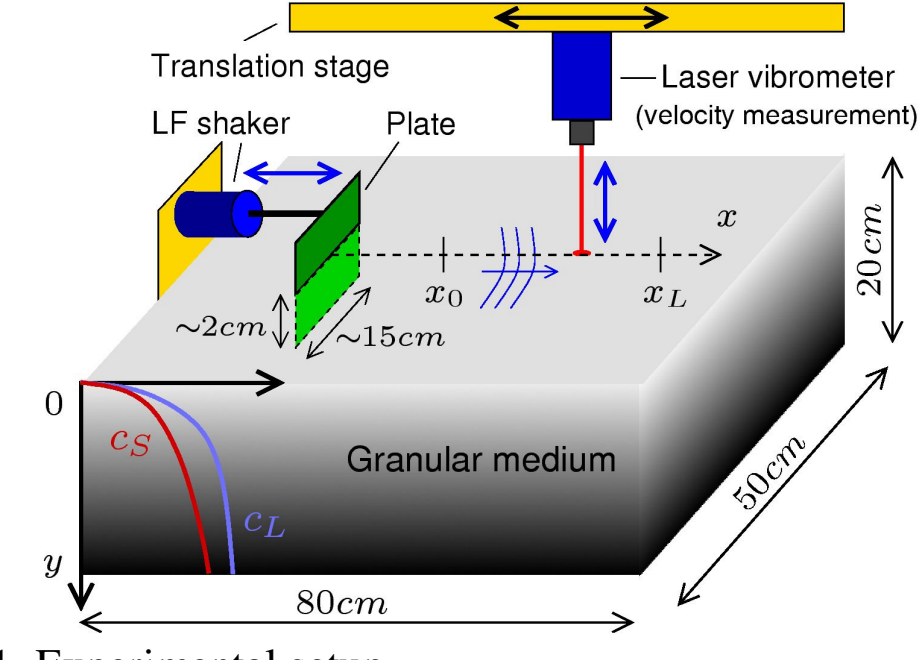

FIG. 1. Experimental setup.

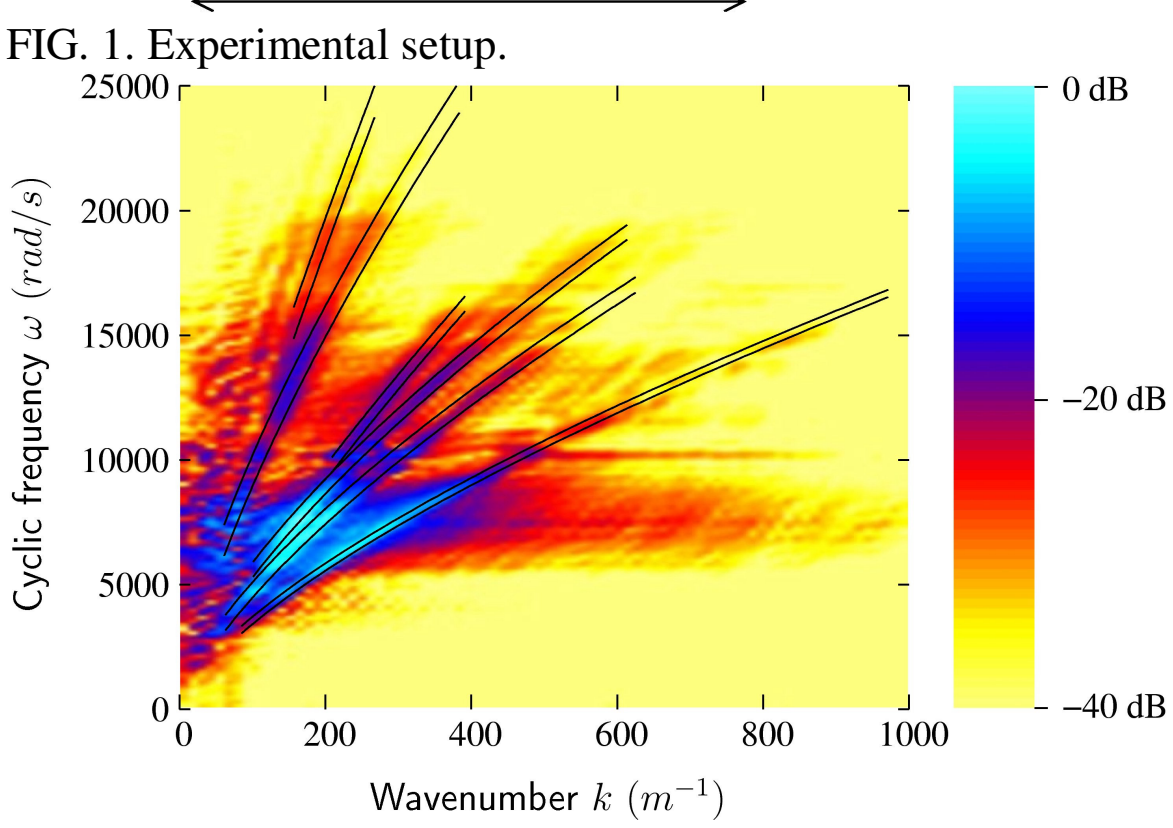

FIG. 2. A typical dispersion curve. Thin lines bounding the "tubes" indicate average + standard deviation and average - standard deviation curves obtained as a result of a statistical treatment of a number of patterns.

For the interpretation of the experimental data, guided surface acoustic modes (GSAMs) [18,19] polarized in the sagittal (vertical) plane are modeled with the continuous medium approximation assuming local isotropy of the elastic properties. The horizontal $u_x$ and the vertical $u_y$ components of the mechanical displacement vector $\vec{U} = \{u_x(y), u_y(y)\} e^{i\omega t - ikx}$ are controlled in microscopically a inhomogeneous medium with vertical stratification $c_{S,L} = c_{S,L}(y)$ by a system of coupled Helmholtz equations [18,19]

$$\left(c_S^2 u_x'\right)' + \left(\omega^2 - c_L^2 k^2\right) u_x = ik\left[\left(c_L^2 - 2c_S^2\right) u_y' + \left(c_S^2 u_y\right)'\right],$$

$$\left(c_L^2 u_y'\right)' + \left(\omega^2 - c_S^2 k^2\right) u_y = ik\left[c_S^2 u_x' + \left(\left(c_L^2 - 2c_S^2\right) u_x\right)'\right].$$

Here the prime denotes the $y$-derivative. In addition to conditions of zero stress tensor at the mechanically free boundary ($y$=0) the eigen-problem for GSAMs includes the conditions of localization ($u_{x,y}(y \to \infty) \to 0$). In unconsolidated granular media, the stratification of sound velocities is controlled by the dependence of the elastic moduli on pressure, which increases linearly with depth because of gravity ($p = \rho g y \propto y$, where $\rho$ is the density and $g$ is the gravitational acceleration).

Assuming $c_{S,L} = \gamma_{S,L}(\rho g y)^{\alpha_{S,L}}$ , where $\gamma_{S,L}$ are depth-independent coefficients, normalizing the depth-coordinate to the inverse wave number $k^{-1}$ ($ky \Rightarrow y$) and using the substitution $iu_y \Rightarrow u_y$ , we transform the governing equations in a form convenient for numerical evaluation:

$$\left(y^{2\alpha_S} u_x'\right)' + \left(\Omega^2 - \delta y^{2\alpha_L}\right) u_x = \left(\delta y^{2\alpha_L} - 2y^{2\alpha_S}\right) u_y' + \left(y^{2\alpha_S} u_y\right)', \quad (1)$$

$$\delta\left(y^{2\alpha_L} u_y'\right)' + \left(\Omega^2 - y^{2\alpha_S}\right) u_y = -y^{2\alpha_S} u_x' - \left((\delta y^{2\alpha_L} - 2y^{2\alpha_S}) u_x\right)'.$$

Here $\Omega = \omega / (\gamma_S k)(k/\rho g)^{\alpha_S}$ is the normalized frequency and $\delta = (\gamma_L/\gamma_S)^2 (\rho g/k)^{2\alpha_L - 2\alpha_S}$ is a dimensionless parameter equal to the square of the ratio of the longitudinal to shear velocities at a depth of about the GSAM penetration ($y \propto k^{-1}$). In the previously studied case, [18,19] $\alpha_S = \alpha_L \equiv \alpha$ , there was no characteristic spatial scale in the system; the eigen frequencies $\Omega_n$ ($n = 1, 2, \ldots$) of GSAMs and the acoustic field spatial distributions (eigen modes) obtained through the numerical solution [19,20] did not depend on $k$ (because for $\alpha_S = \alpha_L$ the parameter $\delta$ is $k$ -independent). The power-law dispersion relation for GSAMs then follows from the definition of $\Omega$: $\omega_n = \Omega_n \gamma_S (\rho g)^{\alpha} k^{1-\alpha} \propto k^{1-\alpha}$. The solutions lose their self-similarity [18,19] as soon as $\alpha_S \neq \alpha_L$ ; in this case a characteristic spatial scale $y_0$ might be formally defined by $c_S(y_0) = c_L(y_0)$ , and the eigenfunctions of the problem (1) would start to depend on $k$ . However, theoretically the self-similarity should be recovered in the asymptotic limits $\delta << 1$ and $\delta >> 1$ . The former case is impossible in physical reality because $\delta > 1$ due to $c_L \propto \sqrt{B + (4/3)G} > c_S \propto \sqrt{G}$ . For the theoretically predicted relation $\alpha_S > \alpha_L$ the inequality $\delta >> 1$ could be expected for sufficiently short-wavelength GSAMs due to $\delta \propto k^{2(\alpha_S - \alpha_L)}$ . In the asymptotic case $\delta >> 1$ , similar to the situation in which acoustic waves propagate at the interface of the ocean with unconsolidated water-saturated sediments [23], the system (1) can be transformed into equations describing two types of waveguide modes: fast and slow.

The modes, which we call here fast modes (in accordance with the notation of fast seismo-acoustic waves [23]), are purely compressive; their description follows from Eq. (1) when shear rigidity is completely neglected ($c_S = 0$). Correspondingly, the fast modes are controlled by the distribution of the velocity $c_L(y)$ , and the dispersion relation scales as $\omega \propto k^{1-\alpha_L}$ . These modes can be easily detected in experiments with normal laser incidence since they have a significant vertical component of the surface displacement.

However in the same limiting case $c_S \to 0, \delta \to \infty$ there exists a second type of waves termed here the slow modes. They are controlled only by the profile $c_S(y)$ and have a dispersion relation scaling as $\omega \propto k^{1-\alpha_S}$ , but they are not pure shear and contain a compression contribution ($div\vec{U} \propto \delta^{-1}$ ). However, as it has been shown by our numerical analysis of Eq. (1), the corresponding vertical component of the surface displacement is typically very weak except for very low order modes.

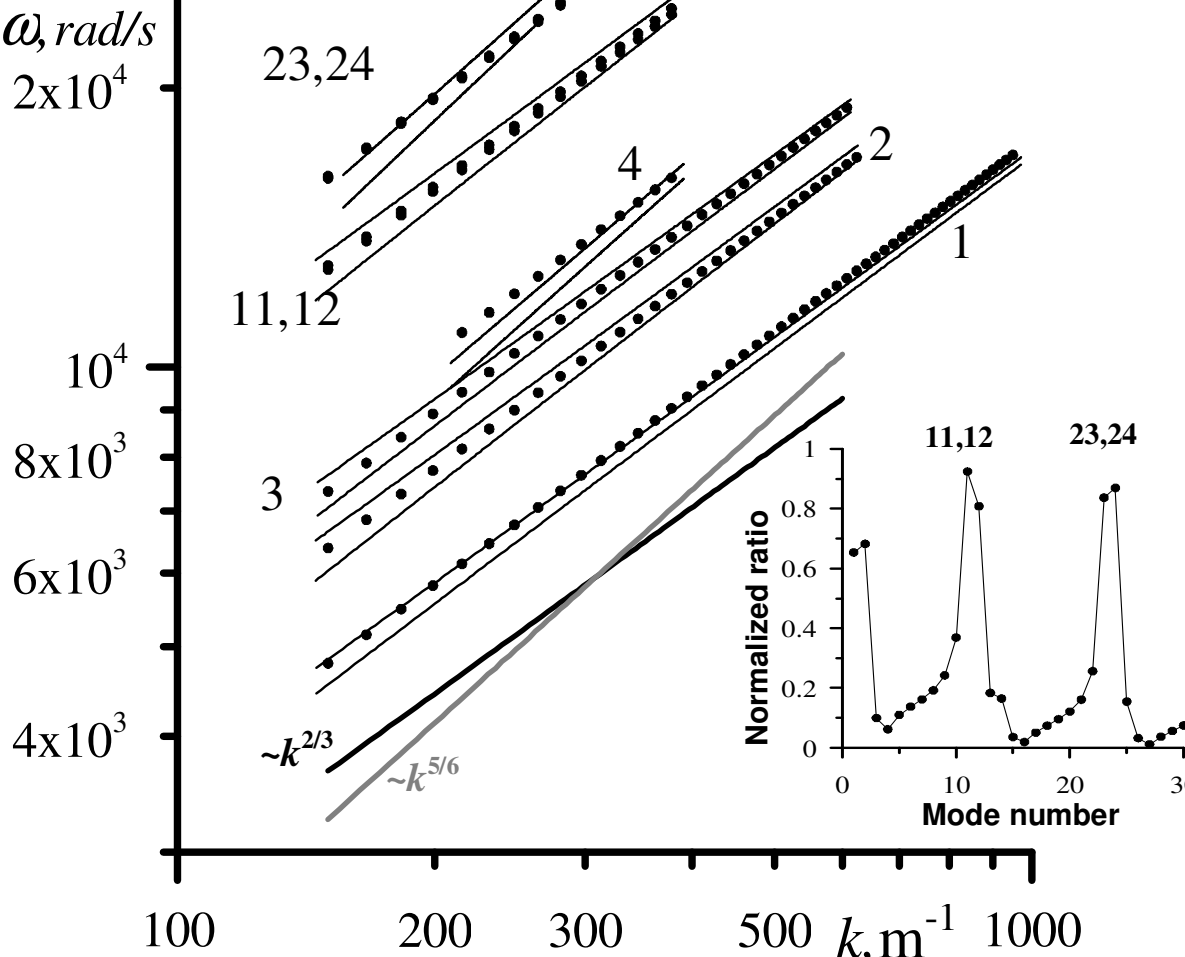

FIG. 3. Experimental dispersion curves (solid lines) represented as ± standard deviation intervals and theoretical points fitted to them. The numbers indicate mode orders. The bunch of solid grey and black lines provides for comparison the theoretical slopes $5/6 = 1 - \alpha_L^{theor}$ and $2/3 = 1 - \alpha_S^{theor}$ . The inset shows a plot of the theoretical normalized ratios of the vertical component of the particle displacement vector at the surface to the modulus of this vector, as a function of mode number.

In the real experimental situation the condition $\delta >> 1$ is fulfilled with a limited precision: our fits (Fig. 3) of the experimental data (Fig. 2) using the numerical solution of Eq. (1) provide $\delta \approx 4.6 - 4.9$ only. Thus, although the experimentally detected modes are neither purely fast nor slow, some of the theoretical dispersion curves for $\delta \approx 4.6 - 4.9$ are mainly determined by $\alpha_S$ and $\gamma_S$ (quasi-slow modes), while the others are mainly determined by the values of $\alpha_L$ and $\gamma_L$ (quasi-fast modes). In this sense, the modes 1-10, 13-22 and 25-30 in our analysis covering the first 30 modes are quasi-slow and thus poorly visible except for some low order modes, whereas the modes 11-12 and 23-24 are quasi-fast and easily detectable. Variations of about 2% in the papameters $\alpha_{S,L}, \gamma_{S,L}$ shift these mode numbers by 1-2. The insert in Fig. 3 shows the theoretical vertical projections of the polarization vector at the surface, i.e. the vertical components of the particle displacement normalized on the displacement modulus, for the modes of different order. This could be used to qualitatively estimate the visibility as measured by the laser vibrometer. We identify the first four lowest "tubes" in Fig. 2 with the first four lowest order (*n* =1-4) GSAMs, and the two other tubes with the four quasi-fast modes 11, 12, 23, 24. The other modes are invisible in experiment (Fig. 2) and are not plotted in Fig. 3.

Matching of the curves is typically realized as the minimization of an objective function, penalizing only points coming out of the corresponding experimental intervals and leaving complete freedom to the points within intervals. For modes higher than the fourth, only points with visibility greater than 0.5 were considered. The minimization was performed by the principal axis method (PRAXIS) developed by Brent [24] and also known as the modified Powell algorithm, which is one of the most effective amongst methods of function optimization without analytical knowledge of derivatives.

This procedure has provided the optimized parameters of the rigidity profiles in Eq. (1): $\gamma_S^{\exp} = 6.43 \pm 0.13$ [SI], $\gamma_L^{\exp} = 14.8 \pm 0.4$ [SI], $\alpha_S^{\exp} = 0.320 \pm 0.006$, and $\alpha_L^{\exp} = 0.305 \pm 0.01$ with the residual value of the objective function around 200 $s^{-1}$. The errors given above have the following sense: if the fitted value is perturbed by the value of the error, the objective function doubles.

Our fitted scaling exponents $\alpha_S^{\exp} \approx 0.96\alpha_S^{theor}$, $\alpha_G^{\exp} = 0.64 \pm 0.01 \approx 0.96\alpha_G^{theor}$ provide the first experimental confirmation for the theoretically predicted behavior of shear rigidity of the 3D disordered packing of spherical elastic beads near the jamming transition $\alpha_G^{theor} = 0.666...$. At the same time, the experimentally revealed scaling exponent of the bulk modulus $\alpha_B^{\exp} = 0.61 \pm 0.02 \approx 1.83\alpha_B^{theor}$ is much closer to the scaling exponent of the shear modulus than to theoretical value $\alpha_B^{theor} = 0.333...$ Our attempt to impose the theoretical value $\alpha_L$ and compensate this misfit by matching the other parameters $\gamma_{S,L}$ and $\alpha_S$ led to an important deviation from the experimental data and to a drastic increase in the residual objective function. We have estimated that the wave velocity dispersion due to medium microinhomogeneity at the scale of the grain size and the diffraction of GSAMs caused by the finite dimensions of our acoustic source [both are not taken into account by the model (1)] give a negligible contribution to the experimentally determined dispersion relations in Fig. 2. We also verified by examining the depth profiles of all eigenmodes visible in experiment that their contact with the bottom of the container can be neglected and that the results obtained for the half-space are self-consistent. The experiment therefore indicates that in the vicinity of the jamming transition near the mechanically free surface, relaxation of the granular media via non-affine motion in response to compression could be very similar to that in response to shear loading. This is in contradiction to the theoretical predictions obtained earlier [3,5-10] for macroscopically homogeneous and isotropic media.

The key to understanding of this discrepancy is the realisation that the granular medium in our experiments is both macroscopically inhomogeneous (not only because of the vertical stratification, but also just because of the presence of the surface [25]) and macroscopically anisotropic [17,26] (because of the preferential direction of its loading by the gravity field). Even in continuum mechanics, inhomogeneity — including the surface as a particular case — and anisotropy induce, through the breaking some symmetry constraints, the coupling of compression and shear motion. Acoustic mode conversion $S \Leftrightarrow L$ takes place at each point of the inhomogeneous media. Pure compression and shear waves can be the eigenmodes of homogeneous anisotropic media only along particular directions, otherwise the eigenmodes are quasi-longitudinal and quasi-shear, combining both compressional and shear motion. We suggest that it could be macroscopic inhomogeneity and anisotropy that through breaking some symmetry constraints couple non-affine motions caused by compression and shear loading on the microscale, allowing disordered granular media under local compression and shear to relax in a similar way. To verify this hypothesis the numerical simulations of the type described in Refs. [5-10] need to include the effects of macroscopic inhomogeneity and anisotropy, and the model in Eq. (1) should be extended to the anisotropic case. The rattlers, particles that do not overlap with the other particles [6] and commonly excluded from the analysis of scaling exponents in granular media[6,13], could also (as any other type of defects) couple compression and shear, contributing an extra relaxation pathway for the granular medium under compressional loading. Equal exponents $\alpha_G$ and $\alpha_B$ are in fact expected according to disordered lattice models for rigidity percolation where all elastic moduli obey the same scaling behavior (see [27] and the references therein).

The success in fitting our experimental results using theoretical predictions for the static (equilibrium) shear modulus indicates that the characteristic time scale $\tau_{n-a}$ of the disordered granular packing non-affine relaxation near the jamming transition ($p \leq 7.5$ kPa) does not exceed few microseconds, as is evident from the quasi-equilibrium condition $\omega\tau_{n-a} << 1$. To determine the relaxation time $\tau_{n-a}$ or a distribution of relaxation times, higher acoustic frequencies than those reported here should be used.

This study was supported by ANR project No. NT05-341989.

*Electronic address: vitali.goussev@univ-lemans.fr